\documentclass[a4paper,10pt]{article}
\usepackage{graphicx}
\usepackage{comment}
\usepackage{amssymb}
\usepackage{amsmath}
\usepackage{nicefrac}
\usepackage{graphicx}
\usepackage{dcolumn}
\usepackage{bm}
\usepackage{comment}
\usepackage{multirow}
\usepackage{cite}

\begin{document}

\newcommand{\bin}[2]{\left(\begin{array}{c}\!#1\!\\\!#2\!\end{array}\right)}
\newcommand{\threej}[6]{\left(\begin{array}{ccc}#1 & #2 & #3 \\ #4 & #5 & #6 \end{array}\right)}
\newcommand{\sixj}[6]{\left\{\begin{array}{ccc}#1 & #2 & #3 \\ #4 & #5 & #6 \end{array}\right\}}
\newcommand{\regge}[9]{\left[\begin{array}{ccc}#1 & #2 & #3 \\ #4 & #5 & #6 \\ #7 & #8 & #9 \end{array}\right]}
\newcommand{\La}[6]{\left[\begin{array}{ccc}#1 & #2 & #3 \\ #4 & #5 & #6 \end{array}\right]}
\newcommand{\hj}{\hat{J}}
\newcommand{\hux}{\hat{J}_{1x}}
\newcommand{\hdx}{\hat{J}_{2x}}
\newcommand{\huy}{\hat{J}_{1y}}
\newcommand{\hdy}{\hat{J}_{2y}}
\newcommand{\huz}{\hat{J}_{1z}}
\newcommand{\hdz}{\hat{J}_{2z}}
\newcommand{\hup}{\hat{J}_1^+}
\newcommand{\hum}{\hat{J}_1^-}
\newcommand{\hdp}{\hat{J}_2^+}
\newcommand{\hdm}{\hat{J}_2^-}

\huge

\begin{center}
New sum rules for Wigner $3jm$ symbols: application to expectation values of hydrogenic ions
\end{center}

\vspace{0.5cm}

\large

\begin{center}
Jean-Christophe Pain$^{a,b,}$\footnote{jean-christophe.pain@cea.fr} and Franck Gilleron$^a$
\end{center}

\normalsize

\begin{center}
\it $^a$CEA, DAM, DIF, F-91297 Arpajon, France\\
\it $^b$Universit\'e Paris-Saclay, CEA, Laboratoire Mati\`ere en Conditions Extr\^emes,\\
\it 91680 Bruy\`eres-le-Ch\^atel, France
\end{center}

\vspace{0.5cm}

\begin{abstract}
We present new sum rules for $3jm$ coefficients, which involve, in addition to the usual weighting factor $(2j+1)$ where $j$ is an angular momentum, the quantity $[j(j+1)]^k$ with $k\ge 1$. The sum rules appear for instance in the statistical modeling of rotational spectra within the theory of moments, and enable one to deduce the expectation values of $r^k$ (used in the theory of Stark effect for hydrogenic ions) in parabolic coordinates from the expectation values of $r^k$ in spherical coordinates.
\end{abstract}

\section{Introduction}\label{sec1} 

Several special relations (identities, sum rules) involving Clebsch-Gordan coefficients or Wigner $3jm$ symbols have been discovered in connection with atomic, molecular and nuclear spectroscopy (see the non-exhaustive list of references \cite{ancarani93,dunlap75,morgan76,rashid76,vandenberghe76,morgan77,demeyer78,klarsfeld78,kulesza80,labarthe80,gazeau80,din81,norcross82,askey82,elbaz85,brudno85,kancerevicius90,ginocchio91,raynal93,ancarani94,minnaert94,casini97,pain12}). Applications concern for instance the hydrogen molecular ion \cite{dunlap75}, the non-relativistic helium atom \cite{morgan76,rashid76,vandenberghe76,morgan77,demeyer78}, the high-order radiative transitions in hydrogenic ions \cite{klarsfeld78}, the calculation of electron-impact cross-sections \cite{ancarani94}, the stability properties of some special classical solutions of the $O(n)$ non-linear $\sigma$-model in two dimensions \cite{din81,askey82}, the non-trivial zeros of $3jm$ and $3nj$ coefficients \cite{brudno85,raynal93}, the pion double charge exchange cross-sections in the nuclear shell model \cite{ginocchio91}, collisions with molecules \cite{norcross82}, the Stark effect of hydrogenic systems \cite{casini97,gilleron19}, or the statistical modeling of anomalous Zeeman patterns \cite{pain12}. Sum rules can be of great interest for checking numerical calculations involving Clebsch-Gordan or Wigner $3jm$ symbols. Most of the published sum rules are gathered in the reference book of Varshalovich, Moskalev and Khersonskii \cite{varshalovich88}. In the present work we provide new sum rules for $3jm$ coefficients, which involve, in addition to the usual weighting factor $(2j+1)$, quantities $[j(j+1)]^k$ with $k\ge 1$ (the case $k=0$ corresponds to the usual orthogonality relation for $3jm$ coefficients):

\begin{equation}
\sum_{j=|j_1-j_2|}^{j_1+j_2}\sum_{m=-j}^j(2j+1)\left[j(j+1)\right]^k\threej{j_1}{j_2}{j}{m_1}{m_2}{-m}^2.
\end{equation}

Such quantities constitute, replacing $j$ by total angular momentum $J$, the angular part of the strength-weighted moments of rotational transition energies \cite{shimamura82,shimamura82b}, which are of interest for the statistical modeling and interpretation of molecular spectra. Some expressions were provided by Ancarani \cite{ancarani93} for $k=1$ to 4 in the case where all projections $m_1$, $m_2$ and $m$ are equal to zero. An algorithm is presented in Sec. \ref{sec2} to obtain such identities for any value of $k$ and any values of the angular momenta and their respective projections. The approach is based on the calculation of powers of the sum of three angular-momentum operators and the use of ladder (or creation / annihilation) operators. Assuming $\hat{\mathbf{J}}=\hat{\mathbf{J}}_1+\hat{\mathbf{J}}_2$, it consists in writing $\hj^2=\hat{A}+\hat{B}_1+\hat{B}_2$ with $\hat{A}=\hj_1^2+\hj_2^2+2\hj_{1z}\hj_{2z}$, $\hat{B}_1=\hat{J}_1^+\hat{J}_2^-$ and $\hat{B}_2=\hat{J}_1^-\hat{J}_2^+$ (with the ladder operators $\hj_i^{\pm}=\hj_{ix}\pm i\hj_{iy}$) and to calculate $\langle j_1,m_1,j_2,m_2|\left(\hat{J}^2\right)^k|j_1,m_1,j_2,m_2\rangle$, retaining in $\hj^2$ only the terms in which the number of occurrences of $\hat{B}_1$ is equal to the number of occurrences of $\hat{B}_2$, in order to keep $m_1$ and $m_2$ unchanged (diagonal matrix elements). Sum rules with $j(j+1)$, $[j(j+1)]^2$ and $[j(j+1)]^3$ respectively are explicitly given. Alternative derivations with weighting factors $j(j+1)$ ($k=1$) and $[j(j+1)]^2$ ($k=2$) for the cases where one projection is equal to zero are discussed in Appendixes A and B respectively. It is outlined, however, that the two latter calculations can not be generalized easily due to the complexity of the method for increasing powers of $j(j+1)$, on the contrary to the algorithm discussed in Sec. \ref{sec2}. Parabolic coordinates represent good quantum numbers for a hydrogenic ion in an external electric field (neglecting fine-structure effects and the mixing of states with different principal quantum numbers). As shown in section \ref{sec3}, the sum rules with one projection equal to zero can be used to deduce the expectation values of $r^k$ (important for instance in the theory of Stark effect for hydrogenic ions) in parabolic coordinates from the expectation values of $r^k$ in spherical coordinates. 

\section{Derivation of the new sum rule}\label{sec2}

\subsection{Angular momentum and ladder operators}\label{subsec21}

Let us consider two angular-momentum operators $\hat{\mathbf{J}}_1$ and $\hat{\mathbf{J}}_2$ such as 

\begin{equation}
\hat{J}_i^2|j_i,m_i\rangle=\hbar^2j_i(j_i+1)|j_i,m_i\rangle\;\;\;\;\mathrm{and}\;\;\;\;\hat{J}_{iz}|j_i,m_i\rangle=m_i\hbar|j_i,m_i\rangle
\end{equation}

\noindent for $i=$1 or 2. In the following, we set $\hbar=1$ for simplicity. If $\hat{\mathbf{J}}=\hat{\mathbf{J}}_1+\hat{\mathbf{J}}_2$ with $\hat{J}^2|j,m\rangle=j(j+1)|j,m\rangle\;\;\;\;\mathrm{and}\;\;\;\;\hat{J}_{z}|j,m\rangle=m|j,m\rangle$, the relation between coupled and uncoupled basis states reads

\begin{equation}
|j_1,m_1,j_2,m_2\rangle=\sum_{j=|j_1-j_2|}^{j_1+j_2}\sum_{m=-j}^j\langle jm|j_1m_1j_2m_2\rangle|j,m\rangle,
\end{equation}

\noindent where $\langle jm|j_1m_1j_2m_2\rangle$ represents the Clebsch-Gordan coefficient 

\begin{equation}
\langle jm|j_1m_1j_2m_2\rangle=(-1)^{j_1-j_2+m}(2j+1)^{1/2}\threej{j_1}{j_2}{j}{m_1}{m_2}{-m}.
\end{equation}

\noindent Let us introduce the so-called ladder operators

\begin{equation}
\begin{array}{l}
\hup=\hux+i\huy\\
\hum=\hux-i\huy\\
\hdp=\hdx+i\hdy\\
\hdm=\hdx-i\hdy.
\end{array}
\end{equation}

\noindent We thus have $\mathbf{J_1.J_2}=\hux\hdx+\huy\hdy+\huz\hdz$, 

\begin{equation}
\hux\hdx+\huy\hdy=\frac{1}{2}\left(\hup\hdm+\hum\hdp\right)
\end{equation}

\noindent and the relations

\begin{eqnarray}
\hup|j_1,m_1,j_2,m_2\rangle&=&\sqrt{j_1(j_1+1)-m_1(m_1+1)}~|j_1,m_1+1,j_2,m_2\rangle,
\end{eqnarray}

\begin{eqnarray}
\hum|j_1,m_1,j_2,m_2\rangle&=&\sqrt{j_1(j_1+1)-m_1(m_1-1)}~|j_1,m_1-1,j_2,m_2\rangle,
\end{eqnarray}

\begin{eqnarray}
\hdp|j_1,m_1,j_2,m_2\rangle&=&\sqrt{j_2(j_2+1)-m_2(m_2+1)}~|j_1,m_1,j_2,m_2+1\rangle,
\end{eqnarray}

\noindent together with

\begin{eqnarray}
\hdm|j_1,m_1,j_2,m_2\rangle&=&\sqrt{j_2(j_2+1)-m_2(m_2-1)}~|j_1,m_1,j_2,m_2-1\rangle
\end{eqnarray}

\noindent yielding

\begin{eqnarray}
\hup\hdm|j_1,m_1,j_2,m_2\rangle&=&\sqrt{j_1(j_1+1)-m_1(m_1+1)}\nonumber\\
& &\times\sqrt{j_2(j_2+1)-m_2(m_2-1)}~|j_1,m_1+1,j_2,m_2-1\rangle
\end{eqnarray}

\noindent as well as

\begin{eqnarray}
\hum\hdp|j_1,m_1,j_2,m_2\rangle&=&\sqrt{j_1(j_1+1)-m_1(m_1-1)}\nonumber\\
& &\times\sqrt{j_2(j_2+1)-m_2(m_2+1)}~|j_1,m_1-1,j_2,m_2+1\rangle.
\end{eqnarray}

\noindent The operators $\hj_1^2$, $\hj_2^2$ and $\hj_{1z}\hj_{2z}$ do not change the values of $m_1$ and $m_2$:

\begin{eqnarray}
& &\left(\hj_1^2+\hj_2^2+2\hj_{1z}\hj_{2z}\right)|j_1,m_1,j_2,m_2\rangle\nonumber\\
&=&\left[j_1(j_1+1)+j_2(j_2+1)+2m_1m_2\right]~|j_1,m_1,j_2,m_2\rangle.
\end{eqnarray}

\noindent Let us set $\hat{A}=\hj_1^2+\hj_2^2+2\hj_{1z}\hj_{2z}$, $\hat{B}_1=\hat{J}_1^+\hat{J}_2^-$ and $\hat{B}_2=\hat{J}_1^-\hat{J}_2^+$. 

\subsection{Starting point of the method: matrix elements of $\left(\hat{J}^2\right)^k$}\label{subsecadded}

In order to obtain a sum rule, we have to calculate

\begin{equation}
\langle j_1,m_1,j_2,m_2|\left(\hat{J}^2\right)^k|j_1,m_1,j_2,m_2\rangle=\sum_j|\langle jm|j_1m_1j_2m_2\rangle|^2 [j(j+1)]^k,
\end{equation}

\noindent where $\hat{J}^2=\hat{A}+\hat{B}_1+\hat{B}_2$. The operator

\begin{equation}\label{j2k}
\left(\hat{J}^2\right)^k=\underbrace{\left(\hat{A}+\hat{B}_1+\hat{B}_2\right).\left(\hat{A}+\hat{B}_1+\hat{B}_2\right)\cdots \left(\hat{A}+\hat{B}_1+\hat{B}_2\right)}_{\mathrm{k~times}},
\end{equation}

\noindent is the sum of $3^ k$ operators, each of them containing $p_1$ operators $\hat{A}$, $p_2$ operators $\hat{B}_1$ and $p_3$ operators $\hat{B}_2$ with $p_1+p_2+p_3=k$. Among those terms, we retain only the ones in which the number of occurrences of $\hat{B}_1$ is equal to the number of occurrences of $\hat{B}_2$  (thus $p_2=p_3$), in order to keep $m_1$ and $m_2$ unchanged (diagonal matrix elements).

\subsection{Sum rule with $j(j+1)$}\label{subsec22}

The right-hand side of Eq. (\ref{j2k}) for $k$=1 is the sum of operators $\hat{A}$, $\hat{B}_1$ and $\hat{B}_2$. 
 
\begin{itemize}

\item The only one that actually contributes to the diagonal terms is $\hat{A}$ and its contribution is $j_1(j_1+1)+j_2(j_2+1)+2m_1m_2$.

\end{itemize}

\noindent The sum rule is therefore 

\begin{eqnarray}
& &\sum_{j=|j_1-j_2|}^{j_1+j_2}\sum_{m=-j}^j(2j+1)j(j+1)\threej{j_1}{j_2}{j}{m_1}{m_2}{-m}^2\nonumber\\
&=&j_1(j_1+1)+j_2(j_2+1)+2m_1m_2.
\end{eqnarray}

\noindent The latter expression is not given in Ref. \cite{varshalovich88}. In the particular case $j_1=a$, $j_2=c$, $j=\ell$, $m_1=-x$, $m_2=x$ and $m=0$, we obtain

\begin{equation}\label{new10}
\sum_{\ell}(2\ell+1)\ell(\ell+1)\threej{a}{c}{\ell}{-x}{x}{0}^2=a(a+1)+c(c+1)-2x^2,
\end{equation}

\noindent for which we found an alternative proof (see Appendix A).
 
\subsection{Sum rule with $\left[j(j+1)\right]^2$}\label{subsec23}

\noindent The right-hand side of Eq. (\ref{j2k}) for $k$=2 is the sum of operators $\hat{A}\hat{A}$, $\hat{A}\hat{B}_1$, $\hat{A}\hat{B}_2$, $\hat{B}_1\hat{A}$, $\hat{B}_2\hat{A}$, $\hat{B}_1\hat{B}_2$, $\hat{B}_2\hat{B}_1$, $\hat{B}_1\hat{B}_1$ and $\hat{B}_2\hat{B}_2$. The only ones that contribute to the diagonal terms are $\hat{A}\hat{A}$, $\hat{B}_1\hat{B}_2$ and $\hat{B}_2\hat{B}_1$, and

\begin{itemize}

\item The contribution of $\hat{A}\hat{A}$ is

\begin{equation}
\left[j_1(j_1+1)+j_2(j_2+1)+2m_1m_2\right]^2.
\end{equation}

\item The contribution of $\hat{B}_1\hat{B}_2$ is 

\begin{equation}
\left[j_1(j_1+1)-m_1(m_1-1)\right]\left[j_2(j_2+1)-m_2(m_2+1)\right].
\end{equation}

\item The contribution of $\hat{B}_2\hat{B}_1$ is 

\begin{equation}
\left[j_1(j_1+1)-m_1(m_1+1)\right]\left[j_2(j_2+1)-m_2(m_2-1)\right].
\end{equation}

\end{itemize}

\noindent The sum rule is therefore

\begin{eqnarray}
& &\sum_{j=|j_1-j_2|}^{j_1+j_2}\sum_{m=-j}^j(2j+1)j^2(j+1)^2\threej{j_1}{j_2}{j}{m_1}{m_2}{-m}^2\nonumber\\
&=&\left[j_1(j_1+1)+j_2(j_2+1)+2m_1m_2\right]^2\nonumber\\
& &+\left[j_1(j_1+1)-m_1(m_1-1)\right]\left[j_2(j_2+1)-m_2(m_2+1)\right]\nonumber\\
& &+\left[j_1(j_1+1)-m_1(m_1+1)\right]\left[j_2(j_2+1)-m_2(m_2-1)\right].
\end{eqnarray}

\noindent In the particular case where $j_1=a$, $j_2=c$, $j=\ell$, $m_1=-x$, $m_2=x$ and $m=0$, we get

\begin{eqnarray}\label{new20}
& &\sum_{\ell}(2\ell+1)\ell^2(\ell+1)^2\threej{a}{c}{\ell}{x}{-x}{0}^2=\left[a(a+1)+c(c+1)-2x^2\right]^2\nonumber\\
& &+\left[a(a+1)-x(x-1)\right]\left[c(c+1)-x(x-1)\right]\nonumber\\
& &+\left[a(a+1)-x(x+1)\right]\left[c(c+1)-x(x+1)\right],
\end{eqnarray}

\noindent which we could obtain in an alternative way as well (see Appendix B), in the same spirit as the $\ell(\ell+1)$ case in Appendix A. However that approach is difficult to generalize to higher powers of $\ell(\ell+1)$, on the contrary to the one presented here.

\subsection{Sum rule with $\left[j(j+1)\right]^3$}\label{subsec24}

The right-hand side of Eq. (\ref{j2k}) in the case $k$=3 can be expanded as the sum of $\hat{A}\hat{A}\hat{A}$, $\hat{A}\hat{A}\hat{B}_1$, $\hat{A}\hat{A}\hat{B}_2$, $\hat{A}\hat{B}_1\hat{A}$, $\hat{A}\hat{B}_2\hat{A}$, $\hat{B}_1\hat{A}\hat{A}$, $\hat{B}_2\hat{A}\hat{A}$, $\hat{A}\hat{B}_1\hat{B}_1$, $\hat{A}\hat{B}_2\hat{B}_2$, $\hat{A}\hat{B}_1\hat{B}_2$, $\hat{A}\hat{B}_2\hat{B}_1$, $\hat{B}_1\hat{B}_1\hat{A}$, $\hat{B}_2\hat{B}_2\hat{A}$, $\hat{B}_1\hat{B}_2\hat{A}$, $\hat{B}_2\hat{B}_1\hat{A}$, $\hat{B}_1\hat{A}\hat{B}_1$, $\hat{B}_2\hat{A}\hat{B}_2$, $\hat{B}_1\hat{A}\hat{B}_2$, $\hat{B}_2\hat{A}\hat{B}_1$, $\hat{B}_1\hat{B}_1\hat{B}_1$, $\hat{B}_1\hat{B}_2\hat{B}_2$, $\hat{B}_1\hat{B}_1\hat{B}_2$, $\hat{B}_1\hat{B}_2\hat{B}_1$, $\hat{B}_2\hat{B}_1\hat{B}_1$, $\hat{B}_2\hat{B}_2\hat{B}_2$, $\hat{B}_2\hat{B}_1\hat{B}_2$ and $\hat{B}_2\hat{B}_2\hat{B}_1$. The only ones that contribute the the diagonal terms are $\hat{A}\hat{A}\hat{A}$, $\hat{A}\hat{B}_1\hat{B}_2$, $\hat{A}\hat{B}_2\hat{B}_1$, $\hat{B}_1\hat{B}_2\hat{A}$, $\hat{B}_2\hat{B}_1\hat{A}$, $\hat{B}_1\hat{A}\hat{B}_2$ and $\hat{B}_2\hat{A}\hat{B}_1$. Since $\left[\hat{A},\hat{B}_1\hat{B}_2\right]$=0, the contributions of $\hat{A}\hat{B}_1\hat{B}_2$ and $\hat{B}_1\hat{B}_2\hat{A}$ are identical. In the same way, since $\left[\hat{A},\hat{B}_2\hat{B}_1\right]$=0, the contributions of $\hat{A}\hat{B}_2\hat{B}_1$ and $\hat{B}_2\hat{B}_1\hat{A}$ are identical as well. Finally, one has

\begin{itemize}

\item The contribution of $\hat{A}\hat{A}\hat{A}$ is 

\begin{equation}
\left[j_1(j_1+1)+j_2(j_2+1)+2m_1m_2\right]^3.
\end{equation}

\item The contribution of $\hat{B}_1\hat{B}_2\hat{A}$, as well as of $\hat{A}\hat{B}_1\hat{B}_2$, is 

\begin{eqnarray}
& &\left[j_1(j_1+1)-m_1(m_1-1)\right]\left[j_1(j_1+1)+j_2(j_2+1)+2m_1m_2\right]\nonumber\\
& &\times\left[j_2(j_2+1)-m_2(m_2+1)\right].
\end{eqnarray}

\item The contribution of $\hat{B}_1\hat{A}\hat{B}_2$ is 

\begin{eqnarray}
& &\left[j_1(j_1+1)-m_1(m_1-1)\right]\left[j_1(j_1+1)+j_2(j_2+1)+2(m_1-1)(m_2+1)\right]\nonumber\\
& &\times\left[j_2(j_2+1)-m_2(m_2+1)\right].
\end{eqnarray}

\item The contribution of $\hat{B}_2\hat{B}_1\hat{A}$, as well as of $\hat{A}\hat{B}_2\hat{B}_1$, is 

\begin{eqnarray}
& &\left[j_1(j_1+1)-m_1(m_1+1)\right]\left[j_1(j_1+1)+j_2(j_2+1)+2m_1m_2\right]\nonumber\\
& &\times\left[j_2(j_2+1)-m_2(m_2-1)\right].
\end{eqnarray}

\item The contribution of $\hat{B}_2\hat{A}\hat{B}_1$ is 

\begin{eqnarray}
& &\left[j_1(j_1+1)-m_1(m_1+1)\right]\left[j_1(j_1+1)+j_2(j_2+1)+2(m_1+1)(m_2-1)\right]\nonumber\\
& &\left[j_2(j_2+1)-m_2(m_2-1)\right].
\end{eqnarray}

\end{itemize}

\noindent The sum rule is therefore

\begin{eqnarray}
& &\sum_{j=|j_1-j_2|}^{j_1+j_2}\sum_{m=-j}^j(2j+1)j^3(j+1)^3\threej{j_1}{j_2}{j}{m_1}{m_2}{-m}^2\nonumber\\
&=&\left[j_1(j_1+1)+j_2(j_2+1)+2m_1m_2\right]^3\nonumber\\
& &+\left[j_1(j_1+1)-m_1(m_1+1)\right]\left[j_2(j_2+1)-m_2(m_2-1)\right]\nonumber\\
& &\times \left[j_1(j_1+1)+j_2(j_2+1)+2(m_1+1)(m_2-1)\right]\nonumber\\
& &+2\left[j_1(j_1+1)-m_1(m_1+1)\right]\left[j_2(j_2+1)-m_2(m_2-1)\right]\nonumber\\
& &\times\left[j_1(j_1+1)+j_2(j_2+1)+2m_1m_2\right]\nonumber\\
& &+2\left[j_1(j_1+1)-m_1(m_1-1)\right]\left[j_2(j_2+1)-m_2(m_2+1)\right]\nonumber\\
& &\times\left[j_1(j_1+1)+j_2(j_2+1)+2m_1m_2\right]\nonumber\\
& &+\left[j_1(j_1+1)-m_1(m_1-1)\right]\left[j_2(j_2+1)-m_2(m_2+1)\right]\nonumber\\
& &\times\left[j_1(j_1+1)+j_2(j_2+1)+2(m_1-1)(m_2+1)\right],
\end{eqnarray}

\noindent and setting $j_1=a$, $j_2=c$, $j=\ell$, $m_1=-x$, $m_2=x$ and $m=0$, we get

\begin{eqnarray}
& &\sum_{\ell}(2\ell+1)\ell^3(\ell+1)^3\threej{a}{c}{\ell}{-x}{x}{0}^2=\left[a(a+1)+c(c+1)-2x^2\right]^3\nonumber\\
& &+\left[a(a+1)-x(x+1)\right]\left[c(c+1)-x(x+1)\right]\nonumber\\
& &\times\left[a(a+1)+c(c+1)-2(x+1)^2\right]\nonumber\\
& &+2\left[a(a+1)-x(x+1)\right]\left[c(c+1)-x(x+1)\right]\nonumber\\
& &\times\left[a(a+1)+c(c+1)-2x^2\right]\nonumber\\
& &+2\left[a(a+1)-x(x-1)\right]\left[c(c+1)-x(x-1)\right]\nonumber\\
& &\times\left[a(a+1)+c(c+1)-2x^2\right]\nonumber\\
& &+\left[a(a+1)-x(x-1)\right]\left[c(c+1)-x(x-1)\right]\nonumber\\
& &\times\left[a(a+1)+c(c+1)-2(x-1)^2\right].
\end{eqnarray}

\subsection{Generalization to higher powers}\label{subsecaddedbis}

The procedure can be easily generalized to higher powers of $j(j+1)$ using a computer algebra system such as Mathematica \cite{mathematica}:

\begin{verbatim}

SumRule[p_, J1_, M1_, J2_, M2_] := Module [
{sumrule, i, j, k, A, B1, B2, aux, perm, term},

A[j1_, m1_, j2_, m2_, val_] :=
        {j1, m1, j2, m2, val (j1 (j1 + 1) + j2 (j2 + 1)+2 m1 m2)};
B1[j1_, m1_, j2_, m2_, val_] := 
        {j1, m1+1, j2, m2-1, val Sqrt[j1 (j1 + 1) - m1 (m1 + 1)] 
                             Sqrt[j2 (j2 + 1) - m2 (m2 - 1)]};
B2[j1_, m1_, j2_, m2_, val_] := 
        {j1, m1-1, j2, m2+1, val Sqrt[j1 (j1 + 1) - m1 (m1 - 1)] 
                             Sqrt[j2 (j2 + 1) - m2 (m2 + 1)]};
sumrule = 0;
Do [ (* List of p operators: even number k of ladder 
                                 operators B and p-k operators A *)
      aux = Flatten[Table[{B1,B2},{j,1,k/2}]~Join~
            Flatten[Table[A,{j,1,p-k}],1],1];
      (* Generate all possible permutations of these p operators *)
      perm = Permutations[aux];
      (* For each permutation, compute the contribution 
                             of the product of these p operators *)
      Do [ 
         aux=perm[[i]];
         term={J1,M1,J2,M2,1};
         Do[term=aux[[j]]@@term;,{j,Length[aux],1,-1}];
         sumrule += term[[5]];,{i,1,Length[perm]}
         ],{k,0,p,2}
   ];
   sumrule
]
  
SumRule[1, J1, M1, J2, M2]
SumRule[2, J1, M1, J2, M2]
SumRule[3, J1, M1, J2, M2]
...
\end{verbatim}

We did not succeed in finding an explicit form for any value of the power of $j(j+1)$. It is worth mentioning, however, that recent results concerning the binomial expansion of the power of the sum of two non-commuting operators could bring new insights into the problem, at least in order to derive alternative algorithms or recurrence relations (see Appendix C). The sum rule can also be obtained from recurrence relations (see Appendix D). For $j(j+1)$, $\left[j(j+1)\right]^2$ and $\left[j(j+1)\right]^3$, a single recurrence on the power of $j(j+1)$ applies, but unfortunately for powers larger than 4, one has to solve coupled recurrence relations, involving diagonal terms (giving the sum rules discussed in the present work) and non-diagonal terms (which leads to other kinds of sum rules).

\section{Hydrogenic expectation values in spherical and parabolic coordinates} \label{sec3}

\subsection{Spherical coordinates}\label{subsec31}

The expectation value of $r^p$ for a hydrogenic atom of charge $Z$ is equal to (in atomic units):

\begin{equation}
\langle n\ell|r^p|n\ell\rangle=\int_0^{\infty}\left[R_{n\ell}(r)\right]^2r^{p+2}dr,
\end{equation}

\noindent where $R_{n\ell}$ is the radial part of the wavefunction:

\begin{eqnarray}\label{rn}
& &R_{n\ell}(r)=\sqrt{\frac{Z(n-\ell-1)!}{n^2(n+\ell)!}}\left(\frac{2Z}{n}\right)^{\ell+1}r^{\ell}\exp\left[-\frac{rZ}{n}\right]L_{n-\ell-1}^{2\ell+1}\left(\frac{2rZ}{n}\right),
\end{eqnarray}

\noindent $L_a^b(x)$ representing a generalized Laguerre polynomial (sometimes referred to as associated Laguerre polynomial or Sonine polynomial). In the following, we concentrate on positive values of $p$. The expectation value for negative values of the exponent are related to the latter by \cite{marxer91}:

\begin{equation}
\langle n\ell |r^p|n\ell\rangle=\frac{(2\ell+p+2)!}{(2\ell-p-2)!}\left(\frac{n}{2Z}\right)^{2p+3}\langle n\ell|r^{-p-2}|n\ell\rangle
\end{equation}

\noindent and one has

\begin{eqnarray}
\langle n\ell|r^{-p-2}|n\ell\rangle&=&\left(\frac{Z}{n\ell}\right)^{p+2}\frac{\ell}{n}\frac{(2\ell)^{p+1}(2\ell-p)!}{(2\ell+1)!}\nonumber\\
& &\times~_3F_2\left[
\begin{array}{c}
-p,p+1,\ell+1-n\\
1,2\ell+2\\
\end{array};1\right],
\end{eqnarray}

\noindent where $_3F_2$ is a hypergeometric series \cite{duverney20}. Expectation values have been derived by various methods since the beginning of quantum mechanics. The earliest attempts can be found in \cite{pasternack37,pasternack37b}, for more recent work see \cite{drake90,curtis91,marxer91} and the references cited therein. Such quantities can be obtained from the Pasternack recurrence relations \cite{pasternack62}:

\begin{equation}
a_k=\frac{(2k+1)n^2}{Z(k+1)}a_{k-1}-\frac{kn^2}{4Z^2(k+1)}\left[4\ell(\ell+1)+1-k^2\right]a_{k-2},
\end{equation}

\noindent with $a_0=1$ and $a_1=\frac{3n^2-\ell(\ell+1)}{2Z}$ and $a_k=\langle n\ell|r^k|n\ell\rangle$, or by the explicit formula

\begin{eqnarray}
\langle n\ell|r^p|n\ell\rangle &=&\frac{Z(n-\ell-1)!}{n^2(n+\ell)!}\sum_{s=0}^{2n-2\ell-2}\frac{(-1)^s}{s!}\frac{\Gamma(2\ell+p+s+3)}{\left(\frac{2Z}{n}\right)^{p+1}}\nonumber\\
& &\times\sum_{j=0}^s\bin{s}{j}\bin{n+\ell}{2\ell+1+j}\bin{n+\ell}{2\ell+1+s-j},
\end{eqnarray}

\noindent $\Gamma(x)$ being the usual Gamma function and $\bin{n}{p}=n!/p!/(n-p)!$ the binomial coefficient.

\subsection{Parabolic coordinates}\label{subsec32}

For quantum mechanical calculations when a particular direction in space is distinguished by some external force, for instance in the case of Stark effect, it is often more convenient to work in a parabolic $|nqm\rangle$ instead of spherical $|n\ell m\rangle$ basis (see \emph{e.g.} \cite{bethe57,landau,sholin73,demura18}). Indeed, the Hamiltonian $\hat{H}_S=F z$ is diagonal in parabolic coordinates for states within the same shell $n$, which turns out to be useful for the application of perturbation theory to a degenerate system \cite{bethe57,landau,lisitsa94,hey07,hey15}. Moreover, the Schr\"odinger equation for the unperturbed one-electron system can also be separated in parabolic coordinates, a fact related to the accidental degeneracy of level energies with respect to the orbital quantum number $\ell$. Schr\"odinger's equation of a hydrogen-like system in parabolic coordinates $|nqm\rangle$ reads:

\begin{equation}
\hat{H}_0|nqm\rangle=-\frac{Z^2}{2n^2}|nqm\rangle,
\end{equation}

\noindent where

\begin{itemize}

\item $n$ is the principal quantum number,
\item $m$ is the orbital magnetic quantum number ($-n+1\le m \le n-1$),
\item $q=n_1-n_2$ is the parabolic (or electric) quantum number, depending on two positive integers $n_1$ and $n_2$ which obey the equation:
$n=n_1+n_2+|m|+1$.

\end{itemize}

\noindent A parabolic state can be represented equivalently as $|n n_1 n_2 m\rangle$ or $|n q m\rangle$ and the wavefunction in parabolic coordinates can be put in the form (we assume $m>0$):

\begin{equation}
\psi_{n,n_1,n_2,m}(\xi,\eta,\phi)=u_{n,n_1,m}(\xi)v_{n,n_2,m}(\eta)e^{im\phi},
\end{equation}

\noindent where the functions $u$ and $v$ read respectively 

\begin{equation}
u_{n,n_1,m}(\xi)=\frac{(-1)^{n_1-m}}{(\pi n)^{1/4}}\left(\frac{Z}{n}\right)^{\frac{(2m+3)}{4}}\sqrt{\frac{n_1!}{(n_1+m)!}}\xi^{m/2}e^{-\frac{Z\xi}{2n}}L_{n_1}^m\left(\frac{Z\xi}{n}\right)
\end{equation}

\noindent and

\begin{equation}
v_{n,n_2,m}(\xi)=\frac{(-1)^{n_2-m}}{(\pi n)^{1/4}}\left(\frac{Z}{n}\right)^{\frac{(2m+3)}{4}}\sqrt{\frac{n_2!}{(n_2+m)!}}\eta^{m/2}e^{-\frac{Z\eta}{2n}}L_{n_2}^m\left(\frac{Z\eta}{n}\right).
\end{equation}

\noindent The expression of $r$ in terms of variables $\xi$ and $\eta$ is $r=\left(\xi+\eta\right)/2$ and the elementary volume

\begin{equation}
d^3r=r^2\sin\theta drd\theta d\phi=\frac{\left(\xi+\eta\right)}{4}d\xi d\eta d\phi.
\end{equation}

\noindent For a diagonal element, one has $n=n'$, $n_1=n_1'$, $n_2=n_2'$ and $m=m'$ and the expectation value is

\begin{eqnarray}
\langle nn_1n_2m|r^p|nn_1n_2m\rangle &=&\int_0^{\infty}\int_0^{\infty}\int_0^{2\pi}\left(\frac{\xi+\eta}{4}\right)\left(\frac{\xi+\eta}{2}\right)^p\nonumber\\
& &\times|u_{n,n_1,m}(\xi)|^2\times |v_{n,n_2,m}(\eta)|^2d\xi d\eta d\phi,
\end{eqnarray}

\noindent \emph{i.e.}

\begin{eqnarray}
\langle nn_1n_2m|r^p|nn_1n_2m\rangle &=&\frac{\pi}{2^{p+1}}\sum_{r=0}^{p+1}\bin{p+1}{r}\langle nn_1m|\xi^{r}|nn_1m\rangle\nonumber\\
& &\times\langle nn_2m|\eta^{p+1-r|}nn_2m\rangle,
\end{eqnarray}

\noindent where the quantity 

\begin{equation}\label{def}
\langle nn_1m|\xi^k|nn_1m\rangle=\int_0^{\infty}\xi^{k+m}e^{-\frac{Z\xi}{n}}\left[L_{n_1}^m\left(\frac{Z\xi}{n}\right)\right]^2d\xi
\end{equation}

\noindent can be obtained using the Hey recurrence relation, which is the equivalent of Pasternack's recursion (derived in spherical coordinates) in parabolic coordinates \cite{hey07}:

\begin{eqnarray}
& &b_k=\frac{n}{Z}\frac{(2k-1)}{k}(2n_1+m+1)b_{k-1}-\left(\frac{n}{Z}\right)^2\frac{(k-1)(m-k+1)(m+k-1)}{k}b_{k-2},\nonumber\\
& &
\end{eqnarray}

\noindent with $b_0=\frac{\sqrt{Z}}{\sqrt{\pi}n}$, $b_1=\frac{2n_1+m+1}{\sqrt{\pi Z}}$ and $b_k=\langle nn_1m|\xi^{k}|nn_1m\rangle$. Note that our definition of $\langle nn_1m|\xi^{k}|nn_1m\rangle$ differs from Hey's one, in the sense that the integral of the right-hand side of Eq. (\ref{def}) corresponds to $\langle nn_1m|\xi^{k-1}|nn_1m\rangle$ for Hey \cite{hey07}. It is also possible to obtain the following explicit form

\begin{eqnarray}\label{sol}
& &\langle nn_1m|\xi^{k}|nn_1m\rangle=\frac{n_1!(m+k)!}{\sqrt{\pi n}(n_1+m)!}\left(\frac{n}{Z}\right)^{k-1/2}\sum_{i=0}^{n_1}\bin{n_1-i-k-1}{n_1-i}^2\bin{m+k+i}{i},\nonumber\\
& &
\end{eqnarray}

\noindent where the binomial coefficient must be understood as an analytical extension of the usual binomial coefficient for negative values of the larger argument, \emph{i.e.}:

\begin{equation}
\bin{g}{p}=\left\{
\begin{array}{ll}
\frac{g!}{(g-p)!p!} & \mathrm{if} \;\;\;\; g>p\geq 0 \\
(-1)^p\frac{(p-g-1)!}{p!(-g-1)!} & \mathrm{if} \;\;\;\; g<0 \;\;\;\; \mathrm{and} \;\;\;\; p\geq 0 \\
0 & \mathrm{if} \;\;\;\; p<0.
\end{array}
\right.
\end{equation}

\begin{table}[]
\centering
\begin{tabular}{|c|c|}\hline
 Order & $\langle n\ell |r^k|n\ell\rangle$ \\
 $k$   & (spherical) \\\hline
 1 & $\frac{1}{2Z}\left[3n^2-\ell(\ell+1)\right]$ \\
 2 & $\frac{n^2}{2Z^2}\left[5n^2+1-3\ell(\ell+1)\right]$ \\
 3 & $\frac{n^2}{8Z^3}\left\{3\ell(\ell+1)\left[\ell(\ell+1)-2\right]+5n^2\left[5-6\ell(\ell+1)\right]+35n^4\right\}$ \\
 4 & $\frac{n^4}{8Z^4}\left\{12-5\ell(\ell+1)\left[10-3\ell(\ell+1)\right]+35n^2\left[3-2\ell(\ell+1)\right]+63n^4\right\}$ \\\hline
\end{tabular}
\caption{Expectation values $\langle n\ell |r^k|n\ell\rangle$ for $k$=1, 2, 3 and 4.}\label{tab1}
\end{table}

\begin{table}[]
\centering
\begin{tabular}{|c|c|}\hline
 Order & $\langle nqm|r^k|nqm\rangle$\\
 $k$   & (parabolic)\\\hline
 1 & $\frac{1}{2Z}\left[3n^2-\frac{1}{2}\left(n^2-1+m^2-q^2\right)\right]$ \\
 2 & $\frac{n^2}{2Z^2}\left[n(2n+3)+\frac{3}{2}\left\{(n-1)^2+q^2-m^2\right\}+1\right]$ \\
 3 & $\frac{n^2}{64Z^3}\left\{33+9m^4+169n^4-6m^2\left[7+19n^2+3q^2\right]\right.$ \\
  & $\left.+2n^2\left[139+51q^2\right]+9\left[6+q^2\right]q^2\right]$ \\
 4 & $\frac{n^4}{64Z^3}\left\{341+45m^4+269n^4-10m^2\left[29+25n^2+9q^2\right]\right.$ \\
  & $\left.+10n^2\left[83+19q^2\right]+350q^2+45q^4\right]$ \\\hline
\end{tabular}
\caption{Expectation values $\langle nqm|r^k|nqm\rangle$ for $k$=1, 2, 3 and 4.}\label{tab2}
\end{table}

\noindent Integrals of the kind $\langle nn_2m|\eta^{k}|nn_2m\rangle$ are equal to the right-hand side of Eq. (\ref{sol}) where $n_1$ has been replaced by $n_2$.

\subsection{Application of the sum rules: connection between expectation values $\langle n\ell |r^k|n\ell\rangle$ and $\langle nqm|r^k|nqm\rangle$}\label{subsec33}

The parabolic and spherical basis are connected by Clebsch-Gordan coefficients determining a transformation from spherical wavefunctions to parabolic ones. From the classical point of view, the squared Clebsch-Gordan is a joint probability of appearance of specific values of spherical and parabolic quantum numbers. The probability can be obtained by two ways: by pure classical calculations of appearance of a specific values of the parabolic quantum numbers in the angular momenta distribution in a Coulomb field by determining a portion of the electron phase space where conservation conditions for the electron motion are fulfilled or by applying the asymptotic (semi-classical) representation of Clebsch-Gordan coefficients \cite{bureyeva01,bureyeva02,brussaard57}.

Hey noticed that $\langle nqm|r|nqm\rangle$ and $\langle nqm|r^2|nqm\rangle$ can be deduced from $\langle n\ell |r|n\ell\rangle$ and $\langle n\ell |r^2|n\ell\rangle$ respectively by making the replacement \cite{hey07}:

\begin{equation}\label{repla1}
\ell(\ell+1)\rightarrow (m+1)(n-1)+2n_1n_2.
\end{equation}

\noindent One has

\begin{equation}
\langle n\ell m|r^p|n\ell m\rangle=\int_{V}\psi_{n\ell m}^*(r)r^p\psi_{n\ell m}(r)d^3r,
\end{equation}

\noindent where $d^3r=r^2dr\sin\theta d\theta d\phi$ and $\psi_{n\ell m}(r,\theta,\phi)=R_{n\ell}(r)Y_{\ell}^m(\theta,\phi)$. The spherical harmonics $Y_{\ell}^{m}(\theta,\phi)$ satisfy the orthogonality relation

\begin{equation}
\int_{\theta=0}^{\pi}\int_{\phi=0}^{2\pi}\left[Y_{\ell}^{m}(\theta,\phi)\right]^*Y_{\ell'}^{m'}(\theta,\phi)\sin\theta d\theta d\phi=\delta_{\ell\ell'}\delta_{mm'}
\end{equation}

\noindent and we have

\begin{eqnarray}
\langle n\ell m|r^p|n\ell m\rangle &=&\int_{\theta=0}^{\pi}\int_{\phi=0}^{2\pi}\left[Y_{\ell}^{m}(\theta,\phi)\right]^*Y_{\ell}^{m}(\theta,\phi)\sin\theta d\theta d\phi\nonumber\\
& &\times\int_0^{\infty}\left[R_{n\ell}(r)\right]^2r^{p+2}dr\nonumber\\
&=&\langle n\ell|r^p|n\ell\rangle.
\end{eqnarray}

\noindent The change of basis from spherical to parabolic coordinates implies that the expectation value in the latter coordinate system reads:

\begin{equation}
\langle nqm|r^p|nqm\rangle=\sum_{\ell}|\langle nqm|n\ell m\rangle|^2\langle n\ell m|r^p|n\ell m\rangle.
\end{equation}

\noindent Therefore, since $\langle n\ell|r^p|n\ell\rangle$ can be expressed as a polynomial of $\ell(\ell+1)$:

\begin{equation}
\langle n\ell|r^p|n\ell\rangle=\sum_{i=0}^pd_{i,p}\left[\ell(\ell+1)\right]^i,
\end{equation}

\noindent then $\langle nqm|r^p|nqm\rangle$ can be expressed as 

\begin{equation}
\langle nqm|r^p|nqm\rangle=\sum_{i=0}^pd_{i,p}\mathcal{M}_i,
\end{equation}

\noindent where $\mathcal{M}_i$ is the transformation of $\left[\ell(\ell+1)\right]^i$ in the basis change from spherical to parabolic coordinates:

\begin{equation}
\mathcal{M}_i=\sum_{\ell}|\langle nqm|n\ell m\rangle|^2\left[\ell(\ell+1)\right]^i.
\end{equation}

\noindent We do not provide a rigorous proof of that observation here, but just point out a connection with the average value of angular-momentum operator $\hat{L}^2$: $\langle n\ell|\hat{L}^2|n\ell\rangle$. In fact, $\ell(\ell+1)$ is nothing else than the average value of square-angular-momentum operator $\hat{L}^2$ in spherical coordinates. In order to calculate the average value of $\hat{L}^2$ in parabolic coordinates, one can use the relation:

\begin{equation}
\langle nqm|n\ell m\rangle=(-1)^{\frac{1+m-q-n}{2}}\sqrt{2\ell+1}\threej{\frac{n-1}{2}}{\frac{n-1}{2}}{\ell}{\frac{m-q}{2}}{\frac{m+q}{2}}{-m}.
\end{equation}

\noindent Since 

\begin{equation}
\threej{\frac{n-1}{2}}{\frac{n-1}{2}}{\ell}{\frac{m-q}{2}}{\frac{m+q}{2}}{-m}=\threej{\frac{n-1+m}{2}}{\frac{n-1-m}{2}}{\ell}{-\frac{q}{2}}{\frac{q}{2}}{0}
\end{equation}

\noindent and applying our new sum rule (\ref{new10}) for $a=(n-1+m)/2$, $c=(n-1-m)/2$ and $x=-q/2$, we get

\begin{eqnarray}
\langle nqm|\hat{L}^2|nqm\rangle&=&\sum_{\ell}|\langle n\ell m|nqm\rangle|^2\langle n\ell m|\hat{L}^2|n\ell m\rangle\nonumber\\
&=&\sum_{\ell}(2\ell+1)\threej{\frac{n-1+m}{2}}{\frac{n-1-m}{2}}{\ell}{-\frac{q}{2}}{\frac{q}{2}}{0}^2\ell(\ell+1)\nonumber\\
&=&\frac{1}{2}\left(n^2+m^2-q^2-1\right)=(m+1)(n-1)+2n_1n_2,
\end{eqnarray}

\noindent since one has $2n_1=n-1-m+q$ and $2n_2=n-1-m-q$. Therefore, the maximum power of $\ell(\ell+1)$ in $\langle n\ell|r|n\ell\rangle$ and $\langle n\ell|r^2|n\ell\rangle$ being 1 (see table \ref{tab1}), it is possible to obtain $\langle nqm|r|nqm\rangle$ and $\langle nqm|r^2|nqm\rangle$ using the replacement of Eq. (\ref{repla1}) (see table \ref{tab2}). 

In the same way, calculating the average value of $\hat{L}^4$, one finds, using the sum rule (\ref{new20}) of section \ref{subsec23}:

\begin{eqnarray}\label{repla2}
\langle nqm|\hat{L}^4|nqm\rangle&=&\sum_{\ell}|\langle n\ell m|nqm\rangle|^2\langle n\ell m|\hat{L}^4|n\ell m\rangle\nonumber\\
&=&\sum_{\ell}(2\ell+1)\threej{\frac{n-1+m}{2}}{\frac{n-1-m}{2}}{\ell}{-\frac{q}{2}}{\frac{q}{2}}{0}^2\left[\ell(\ell+1)\right]^2\nonumber\\
&=&\frac{1}{8}\left\{3+3m^4+10q^2+2m^2\left[n^2-3\left(1+q^2\right)\right]\right.\nonumber\\
& &\left.+3\left[n^4+q^4-2m^2\left(1+q^2\right)\right]\right\},
\end{eqnarray}

\noindent with $q=n_1-n_2$. Subsequently, since the maximum power of $\ell(\ell+1)$ in $\langle n\ell|r^3|n\ell\rangle$ and $\langle n\ell|r^4|n\ell\rangle$ is equal to 2 (see table \ref{tab1}), it is possible to obtain $\langle nqm|r^3|nqm\rangle$ and $\langle nqm|r^4|nqm\rangle$ replacing $\ell(\ell+1)$ by $(m+1)(n-1)+2n_1n_2$ (see Eq. (\ref{repla1})) and $\left[\ell(\ell+1)\right]^2$ by $\frac{1}{8}\left\{3+3m^4+10q^2+2m^2\left[n^2-3\left(1+q^2\right)\right]+3\left[n^4+q^4-2m^2\left(1+q^2\right)\right]\right\}$ (see Eq. (\ref{repla2})). 

Curtis also derived semi-classical expressions of $\langle n\ell|r^p|n\ell\rangle$ and mentioned that these expressions differ from their quantum-mechanical equivalents only in the eigenvalues that are generated by successive applications of the angular momentum $\hat{L}$ \cite{curtis91}. In these semi-classical expressions, the matrix element of $\hat{L}^n$ is replaced by $\left(\ell+1/2\right)^n$ and one has

\begin{equation}\label{sc}
\langle r^s\rangle=\left(\frac{n^2}{Z}\right)^s\left(\frac{n}{k_o}\right)^{\lambda-s-1}\left(\frac{k_e}{n}\right)^{\lambda}\mathcal{P}_{\lambda}\left(\frac{n}{k_e}\right).
\end{equation}

\noindent Here the quantity $k$ is associated with the angular momentum and is factored in two parts: $k_e$ leads to even powers and $k_o$ to odd or non-contributing powers ($n^{2s}(n/k_o)^{\lambda-s-1}$ reduces to $n^{-3}k_o^{2s+3}$ for $s\leq -2$ and to $n^{2s}$ for $s\geq -1$). In the semi-classical case, these two quantities are equal: $k_e=k_o=\ell+1/2$, where $1/2$ is the contribution from the Maslov index. To obtain the quantum-mechnical result from Eq. (\ref{sc}), Curtis compared it with the exact results derived by Bockasten \cite{bockasten74,bockasten76} and found the required replacements:

\begin{equation}
\begin{array}{l}
k_o^{2t+1}=(2\ell+t+1)!/[(2\ell-t)!2^{2t+1}]\\
k_e^{2t}=\sum_{r=0}^t(-1)^{t+r}C_{\lambda,t,r}(\ell+r)!/(\ell-r)!
\end{array}
\end{equation}

\noindent with (see Ref. \cite{drake90}) $(-1)^{t+r}C_{\lambda,t,r}=d_{t,r}^{(\lambda+2)}$ and

\begin{eqnarray}
d_{t,r}^{(\lambda+2)}&=&\frac{1}{\lambda(2\lambda-2t-1)}\left[(\lambda-2t)(2\lambda-1)d_{t,r}^{(\lambda+1)}+2t(\lambda-1)\right.\nonumber\\
& &\left.\times\left\{d_{t-1,r-1}^{(\lambda)}+[r(r+1)-\lambda(\lambda-2)/4]d_{t-1,r}^{(\lambda)}\right\}\right],
\end{eqnarray}

\noindent initialized by the starting values $d_{0,0}^{(2)}=1$ and $d_{0,0}^{(3)}=1$. 

It is worth mentioning that, as shown by Bureyeva \emph{et al.} \cite{bureyeva02}, for large values of quantum numbers, the joint probability of spherical and parabolic quantum numbers is, in the quasi-classical limit (corresponding to $m<\ell\ll n$):

\begin{equation}
\langle\ell m|\left(\frac{n-1}{2}\right)\left(\frac{m-q}{2}\right)\left(\frac{n-1}{2}\right)\left(\frac{m+q}{2}\right)\rangle\approx\frac{2\ell}{\pi\sqrt{\left(\ell^2-\ell_{\mathrm{min}}^2\right)\left(\ell_{\mathrm{max}}^2-\ell^2\right)}}
\end{equation}

\noindent with

\begin{equation}
\ell_{\mathrm{min}}^2=\frac{\left[(n-1)^2+m^2-q^2\right]}{2}-\frac{1}{2}\left\{\left[(n-1)^2+m^2-q^2\right]^2-4(n-1)^2m^2\right\}^{1/2}
\end{equation}

\noindent and

\begin{equation}
\ell_{\mathrm{max}}^2=\frac{\left[(n-1)^2+m^2-q^2\right]}{2}+\frac{1}{2}\left\{\left[(n-1)^2+m^2-q^2\right]^2-4(n-1)^2m^2\right\}^{1/2}.
\end{equation}

\noindent In the case where $m\ll n$, one has

\begin{equation}
\ell_{\mathrm{min}}^2\approx (n-1)^2m^2\left[(n-1)^2+m^2-q^2\right]^{-1}
\end{equation}

\noindent and

\begin{equation}
\ell_{\mathrm{max}}^2\approx\left[(n-1)^2+m^2-q^2\right]=(n-1)^2\frac{m^2}{\ell_{\mathrm{min}}^2}.
\end{equation}

\noindent For $m$=0, the probability of the appearance of the parabolic quantum number $q$ is given by

\begin{equation}
\frac{2}{\pi\sqrt{(n-1)^2-q^2}}
\end{equation}

\noindent and therefore does not depend on $\ell$ if $\ell\ll n$. 

Note that if we had replaced $\ell(\ell+1)$ by $(m+1)(n-1)+2n_1n_2$ even in $\left[\ell(\ell+1)\right]^2$ in the quantum-mechanical expectation values, the result would have been wrong. In the same way, in order to find the relation between $\langle n\ell|r^p|n\ell\rangle$ and $\langle n\ell|r^{p+1}|n\ell\rangle$ on one hand, and $\langle nqm|r^p|nqm\rangle$ and $\langle nqm|r^{p+1}|nqm\rangle$ on the other hand, one would require a generalization of sum rules (\ref{new10}) and (\ref{new20}) with the weighting factor $\left[\ell(\ell+1)\right]^p$.

\section{Conclusion}\label{sec4}

A new family of sum rules for Clebsch-Gordan coefficients or Wigner $3jm$ symbols was presented, involving weighting factors which are positive powers of $j(j+1)$. An algorithm was described, and the sum rules for $j(j+1)$, $[j(j+1)]^2$ and $[j(j+1)]^3$ were explicitely given. The technique presented here, involving ladder operators, can be applied, with the help of a computer algebra system, whatever the power of $j(j+1)$, although the complexity of the calculations increases with the exponent. In the case where one of the projections is equal to zero, an alternative approach is discussed for powers 1 and 2, since it is likely to bring new ideas and tools for the derivation of further identities. Such simple but non-trivial relations are involved in the change of basis from spherical to parabolic coordinates. In the general case, the sum rules find applications in the statistical modeling of rotational spectra relying on strength-weighted moments. For instance, the fourth-order moment (related to the so-called kurtosis) enables one to quantify the sharpness (or flatness) of a rotational absorption structure. We plan to investigate further sum rules related to non-diagonal matrix elements of powers of the radial coordinates, in a non-relativistic or relativistic framework \cite{pasternack62,epstein67,badawi73,blanchard74,ojha84,shertzer91,moreno91,morales92,sanchez92,lopez95,andrae97,hey06,blaive09,gonzales17}. 

\appendix

\section{Alternative derivation of the sum rule with $\ell(\ell+1)$}\label{sec5} 

The following general identity (Ref. \cite{varshalovich88}, formula (4) p. 453):

\begin{eqnarray}
& &\sum_{q\kappa}(-1)^{q-\kappa}[q]\threej{a}{b}{q}{-\alpha}{-\beta}{\kappa}\threej{q}{a}{b}{-\kappa}{\alpha'}{\beta'}=(-1)^{a+\alpha+b+\beta}\delta_{\alpha\alpha'}\delta_{\beta\beta'},
\end{eqnarray}

\noindent yields, with $q\equiv\ell$ ($\ell$ being an integer), $b=c$, $\alpha=x$ and $\beta=-x$: 

\begin{equation}
\sum_{\ell}(2\ell+1)\threej{a}{c}{\ell}{-x}{x}{0}^2=1.
\end{equation}

\noindent The following identity (Ref. \cite{varshalovich88}, formula (5) p. 453):

\begin{eqnarray}\label{var}
& &\sum_{\kappa}(-1)^{q-\kappa}\threej{a}{b}{q}{\alpha}{\beta}{-\kappa}\threej{q}{d}{c}{\kappa}{\delta}{\gamma}\nonumber\\
&=&(-1)^{2a}\sum_{y\xi}(-1)^{y-\xi}[y]\threej{a}{c}{y}{\alpha}{\gamma}{-\xi}\threej{y}{d}{b}{\xi}{\delta}{\beta}\sixj{b}{d}{y}{c}{a}{q},\nonumber\\
& &
\end{eqnarray}

\noindent applied with $y\equiv\ell$, $q=1$, $\alpha=\delta=x$, $\beta=\gamma=-x$, $b=a$, $d=c$, gives

\begin{eqnarray}
& &\sum_{\ell}(-1)^{\ell}(2\ell+1)\threej{a}{c}{\ell}{-x}{x}{0}^2\sixj{a}{c}{\ell}{c}{a}{1}\nonumber\\
&=&(-1)^{1-2a}\threej{a}{a}{1}{-x}{x}{0}\threej{c}{c}{1}{-x}{x}{0}.
\end{eqnarray}

\noindent Using the fact that (we assume $a\ge 1$ and $c\ge 1$):

\begin{equation}\label{6j}
\sixj{a}{c}{\ell}{c}{a}{1}=(-1)^{a+c+\ell}\frac{\ell(\ell+1)-a(a+1)-c(c+1)}{2\sqrt{a(a+1)(2a+1)c(c+1)(2c+1)}}.
\end{equation}

\noindent and \cite{edmonds57}

\begin{equation}
\threej{a}{a}{1}{x}{-x}{0}=(-1)^{a-x}\frac{x}{\sqrt{a(a+1)(2a+1)}},
\end{equation}

\noindent we obtain the following expression

\begin{equation}\label{new1}
\sum_{\ell}(2\ell+1)\ell(\ell+1)\threej{a}{c}{\ell}{-x}{x}{0}^2=a(a+1)+c(c+1)-2x^2,
\end{equation}

\noindent which is exactly Eq. (\ref{new10}). Following the same procedure, it is possible to obtain a similar relation, but with $\left[\ell(\ell+1)\right]^2$ instead of $\ell(\ell+1)$.

\section{Alternative approach for the sum rule involving $\left[\ell(\ell+1)\right]^2$}\label{sec6}

Using the following relation (Ref. \cite{varshalovich88}, Eq. (10) p. 455):

\begin{eqnarray}\label{newvar}
& &\sum_{\psi\kappa\rho\sigma}(-1)^{p-\psi+q-\kappa+r-\rho+s-\sigma}\threej{p}{a}{q}{\psi}{\alpha}{-\kappa}\threej{q}{b}{r}{\kappa}{\beta}{-\rho}\nonumber\\
& &\times\threej{r}{c}{s}{\rho}{\gamma}{-\sigma}\threej{s}{d}{p}{\sigma}{\delta}{-\psi}\nonumber\\
&=&(-1)^{s-a-d-q}\sum_{y\xi}(-1)^{y-\xi}[y]\threej{a}{y}{d}{\alpha}{-\xi}{\delta}\threej{b}{y}{c}{\beta}{\xi}{\gamma}\nonumber\\
& &\times\sixj{a}{y}{d}{s}{p}{q}\sixj{b}{y}{c}{s}{r}{q},
\end{eqnarray}

\noindent together with the expression of $\ell(\ell+1)$ as a linear function of the $6j$ symbol from Eq. (\ref{6j}), it is possible to obtain the expression of

\begin{equation}\label{enig}
\sum_{\ell}(2\ell+1)\ell^2(\ell+1)^2\threej{a}{c}{\ell}{x}{-x}{0}^2.
\end{equation}

\noindent However, such an approach, which cornerstone is relation (\ref{newvar}), is rather cumbersome, and although they can be derived using Yutsis' graphical method \cite{yutsis62}, sum rules such as (\ref{var}) and (\ref{newvar}) but with three, four, etc. $6j$ symbols in the summation of the right-hand side are not given in the handbook of Varshalovich \emph{et al.}. Therefore, it would be difficult to extend the procedure to higher powers of $\ell(\ell+1)$.

In fact, the expression for the sum of Eq. (\ref{enig}) can also be derived using expression (\ref{var}) \cite{heim09} for $y=\ell$, $\xi=0$, $q=2$, $\alpha=\delta=x$, $\beta=\gamma=-x$, $b=a$, $d=c$ and

\begin{eqnarray}
\threej{a}{a}{2}{x}{-x}{0}&=&(-1)^{a-x+1}\frac{a(a+1)-3x^2}{\sqrt{a(a+1)(2a-1)(2a+1)(2a+3)}},\nonumber\\
& &
\end{eqnarray}

\noindent as well as \cite{edmonds57}:

\begin{equation}
\sixj{a}{c}{\ell}{c}{a}{2}=2(-1)^{a+b+c}J(a,c,\ell)\sqrt{\frac{(2a-2)!(2c-2)!}{(2a+3)!(2c+3)!}},
\end{equation}

\noindent where $J(a,c,\ell)=3\left[a(a+1)+c(c+1)-\ell(\ell+1)\right]\left[a(a+1)+c(c+1)-\ell(\ell+1)-1\right]-4a(a+1)c(c+1)$. This leads to the following sum rule,

\begin{eqnarray}
& &\sum_{\ell}(2\ell+1)\ell^2(\ell+1)^2\threej{a}{c}{\ell}{x}{-x}{0}^2=\nonumber\\
& &\left[a(a+1)+c(c+1)-2x^2\right]\left[2a(a+1)+2c(c+1)-1\right]\nonumber\\
& &-\left[a(a+1)+c(c+1)\right]\left[a(a+1)+c(c+1)-1\right]\nonumber\\
& &+\frac{4}{3}a(a+1)c(c+1)+\frac{2}{3}\left[a(a+1)-3x^2\right]\left[c(c+1)-3x^2\right],
\end{eqnarray}

\noindent \emph{i.e.}

\begin{eqnarray}\label{new2}
& &\sum_{\ell}(2\ell+1)\ell^2(\ell+1)^2\threej{a}{c}{\ell}{x}{-x}{0}^2=\left[a(a+1)+c(c+1)-2x^2\right]^2\nonumber\\
& &+\left[a(a+1)-x(x-1)\right]\left[c(c+1)-x(x-1)\right]\nonumber\\
& &+\left[a(a+1)-x(x+1)\right]\left[c(c+1)-x(x+1)\right],
\end{eqnarray}

\noindent which is exactly Eq. (\ref{new20}). The same procedure can be generalized to higher powers of $\ell(\ell+1)$, using

\begin{eqnarray}
\sixj{a}{c}{\ell}{c}{a}{3}&=&(-1)^{a+c+\ell+1}4\left\{5[a(a+1)+c(c+1)-\ell(\ell+1)]^3\right.\nonumber\\
& &-20[a(a+1)+c(c+1)-\ell(\ell+1)]^2\nonumber\\
& &-4[a(a+1)+c(c+1)-\ell(\ell+1)]\nonumber\\
& &\times[3a(a+1)c(c+1)-a(a+1)-c(c+1)-3]\nonumber\\
& &\left.+20a(a+1)c(c+1)\right\}\left[\frac{(2a-3)!(2c-3)!}{(2a+4)!(2c+4)!}\right]^{1/2}
\end{eqnarray}

\noindent as well as

\begin{equation}
\threej{a}{a}{3}{x}{-x}{0}=\frac{(-1)^{a-x+1}x\left[3a(a+1)-5x^2-1\right]}{\sqrt{(a-1)a(a+1)(a+2)(2a-1)(2a+1)(2a+3)}}
\end{equation}

\noindent to calculate

\begin{equation}
\sum_{\ell}(2\ell+1)\ell^3(\ell+1)^3\threej{a}{c}{\ell}{x}{-x}{0}^2.
\end{equation}

\noindent More generally, in order to determine 

\begin{eqnarray}
\sum_{\ell}(2\ell+1)\ell^k(\ell+1)^k\threej{a}{c}{\ell}{x}{-x}{0}^2,
\end{eqnarray}

\noindent one possibility is to express $\ell^k(\ell+1)^k$ as

\begin{equation}
\left[\ell(\ell+1)\right]^k=\sum_{i=0}^kb_{k,i}\sixj{a}{a}{i}{c}{c}{\ell}
\end{equation}

\noindent resorting to the recurrence relation (see for instance Ref. \cite{varshalovich88}, Eq. (6) p. 304):

\begin{eqnarray}
(2i+1)\left[-2a(a+1)-2c(c+1)+2\ell(\ell+1)+i(i+1)\right]\sixj{a}{a}{i}{c}{c}{\ell}& &\nonumber\\
\;\;\;\;\;\;\;\;=(i+1)\sqrt{(2a+i+2)(2a-i)(2c+i+2)(2c-i)}\sixj{a}{a}{i+1}{c}{c}{\ell}& &\nonumber\\
\;\;\;\;\;\;\;\;+i\sqrt{(2a+i+1)(2a-i+1)(2c+i+1)(2c-i+1)}\sixj{a}{a}{i-1}{c}{c}{\ell}& &\nonumber\\
& &
\end{eqnarray}

\noindent and therefore the following coefficients are required

\begin{equation}
\sixj{a}{c}{\ell}{c}{a}{i},
\end{equation}

\noindent for $i\leq k$ as well as \cite{edmonds57}

\begin{eqnarray}
\threej{a}{a}{i}{x}{-x}{0}&=&\sqrt{\frac{(2a-i)!}{(2a+i+1)!}}\sum_{n=0}^i\bin{a}{i}^2\left(-a+x\right)_n\left(a+x+1+n-i\right)_{k-n}\nonumber\\
&=&\sqrt{\frac{(2a-i)!}{(2a+i+1)!}}\sum_{n=0}^i(-1)^{i-n}\bin{a}{i}^2\left(-a+x\right)_n\left(-a-x\right)_{i-n},\nonumber\\
& &
\end{eqnarray}

\noindent where $(y)_p=y(y+1)...(y+p-1)$ represents the usual Pochhammer symbol \cite{abramowitz64}. The calculation becomes of course more and more cumbersome as the power $k$ of the weighting factor $\ell(\ell+1)$ increases, but the procedure can be applied using a computer algebra system.

\section{The non-commutative binomial theorem}\label{sec7}

Wyss found the following expression \cite{wyssa}:

\begin{equation}\label{relwyss}
(\hat{A}+\hat{B})^n=\sum_{k=0}^n\bin{n}{k}\left[\hat{A}^k\hat{B}^{n-k}+\mathcal{D}_k(\hat{B},\hat{A})\hat{B}^{n-k}\right],
\end{equation}

\noindent where

\begin{equation}\label{inter}
\mathcal{D}_{k}(\hat{B},\hat{A})=\left[\hat{B},\hat{A}^{k-1}\right]+\hat{A}\mathcal{D}_{k-1}(\hat{B},\hat{A})+\left[\hat{B},\mathcal{D}_{k-1}(\hat{B},\hat{A})\right],
\end{equation}

\noindent initialized with $\mathcal{D}_0(\hat{B},\hat{A})=0$. We have $\hat{B}=\hat{B}_1+\hat{B}_2$. Using

\begin{equation}
\hat{A}^{k-1}|j_1,m_1,j_2,m_2\rangle=\left[j_1(j_1+1)+j_2(j_2+1)+2m_1m_2\right]^{k-1}~|j_1,m_1,j_2,m_2\rangle,
\end{equation}

\noindent we obtain 

\begin{eqnarray}\label{exp1}
\hat{B}_1\hat{A}^{k-1}|j_1,m_1,j_2,m_2\rangle&=&\sqrt{j_1(j_1+1)-m_1(m_1+1)}\sqrt{j_2(j_2+1)-m_2(m_2-1)}\nonumber\\
& &\times\left[j_1(j_1+1)+j_2(j_2+1)+2m_1m_2\right]^{k-1}\nonumber\\
& &~|j_1,m_1+1,j_2,m_2-1\rangle,
\end{eqnarray}

\noindent and

\begin{eqnarray}\label{exp2}
\hat{A}^{k-1}\hat{B}_1|j_1,m_1,j_2,m_2\rangle&=&\sqrt{j_1(j_1+1)-m_1(m_1+1)}\sqrt{j_2(j_2+1)-m_2(m_2-1)}\nonumber\\
& &\times\left[j_1(j_1+1)+j_2(j_2+1)+2(m_1+1)(m_2-1)\right]^{k-1}\nonumber\\
& &~|j_1,m_1+1,j_2,m_2-1\rangle.
\end{eqnarray}

\noindent We have also

\begin{eqnarray}\label{exp3}
\hat{B}_1^{k-1}|j_1,m_1,j_2,m_2\rangle&=&X_{k-1}~|j_1,m_1+k-1,j_2,m_2-k+1\rangle
\end{eqnarray}

\noindent with

\begin{eqnarray}\label{exp4}
X_{k-1}&=&\prod_{p=0}^{k-2}\sqrt{j_1(j_1+1)-(m_1+p)(m_1+p+1)}\sqrt{j_2(j_2+1)-(m_2-p)(m_2-p-1)}.\nonumber\\
& &
\end{eqnarray}

\noindent The latter expressions (\ref{exp1}), (\ref{exp2}), (\ref{exp3}) and (\ref{exp4}) together with their symmetric forms when $\hat{B_1}$ is replaced by $\hat{B}_2$ and \emph{vice versa}, enable one to calculate the commutators $\left[\hat{B},\hat{A}^{k-1}\right]$ (which contribution to the diagonal terms is zero) in Eq. (\ref{inter}) and $\left[\hat{B}_2,\hat{B}_1^{k-1}\right]$ (which contribution is zero if $k\ne 2$). All the other terms involving coefficients $\mathcal{D}_k$ and their commutators are given by recurrence relations. They may help finding recurrence relations on the sum rules themselves, or at least provide alternative algorithms to obtain them.

\section{Recurrence relations}\label{sec8}

Let us set $\hat{\Lambda}_n=\left(\hat{A}+\hat{B}\right)^n$. We can write $\hat{\Lambda}_n=\hat{\Lambda}_n^++\hat{\Lambda}_n^-$, where $\hat{\Lambda}_n^+$ represents the terms of $\hat{\Lambda}_n$ containing an even number of operators $\hat{B}_i$, $i$=1 or 2 and $\hat{\Lambda}_n^-$ an odd number of such operators in the expansion of $\hat{\Lambda}_n$ as a sum. We have

\begin{equation}
\begin{array}{l}
\hat{\Lambda}_{n+1}^+=\hat{\Lambda}_{n}^+\hat{A}+\hat{\Lambda}_{n}^-\hat{B}\\
\hat{\Lambda}_{n+1}^-=\hat{\Lambda}_{n}^+\hat{B}+\hat{\Lambda}_{n}^-\hat{A},
\end{array}
\end{equation}

\noindent initialized by $\hat{\Lambda}_0^+=1$ and $\hat{\Lambda}_0^-=0$. We obtain, by substitution

\begin{equation}
\hat{\Lambda}_{n+1}^+=\hat{\Lambda}_{n}^+\hat{A}+\sum_{k=1}^n\hat{\Lambda}_{n-k}^+\hat{B}\hat{A}^{k-1}\hat{B}.
\end{equation}

\noindent In order to derive our sum rules, we are interested in diagonal terms of $\hat{\Lambda}_{n}^+$. In the present case, we have

\begin{eqnarray}
\langle j_1,m_1,j_2,m_2|\hat{\Lambda}_{n+1}^+|j_1,m_1,j_2,m_2\rangle &=&\langle j_1,m_1,j_2,m_2|\hat{\Lambda}_{n}^+\hat{A}|j_1,m_1,j_2,m_2\rangle\nonumber\\
& &+\sum_{k=1}^n\langle j_1,m_1,j_2,m_2|\hat{\Lambda}_{n-k}^+\hat{B}_1\hat{A}^{k-1}\hat{B}_2|j_1,m_1,j_2,m_2\rangle\nonumber\\
& &+\sum_{k=1}^n\langle j_1,m_1,j_2,m_2|\hat{\Lambda}_{n-k}^+\hat{B}_2\hat{A}^{k-1}\hat{B}_1|j_1,m_1,j_2,m_2\rangle\nonumber\\
& &+\sum_{k=1}^n\langle j_1,m_1,j_2,m_2|\hat{\Lambda}_{n-k}^+\hat{B}_1\hat{A}^{k-1}\hat{B}_1|j_1,m_1,j_2,m_2\rangle\nonumber\\
& &+\sum_{k=1}^n\langle j_1,m_1,j_2,m_2|\hat{\Lambda}_{n-k}^+\hat{B}_2\hat{A}^{k-1}\hat{B}_2|j_1,m_1,j_2,m_2\rangle.\nonumber\\
\end{eqnarray}

\noindent The first-term of the right-hand side gives:

\begin{eqnarray}
\langle j_1,m_1,j_2,m_2|\hat{\Lambda}_{n}^+\hat{A}|j_1,m_1,j_2,m_2\rangle&=&\left(j_1(j_1+1)+j_2(j_2+1)-2m_1m_2\right)\nonumber\\
& &\times\langle j_1,m_1,j_2,m_2|\hat{\Lambda}_{n}^+|j_1,m_1,j_2,m_2\rangle,
\end{eqnarray}

\noindent and the other terms in the summation:

\begin{eqnarray}\label{un}
& &\langle j_1,m_1,j_2,m_2|\hat{\Lambda}_{n-k}^+\hat{B}_1\hat{A}^{k-1}\hat{B}_2|j_1,m_1,j_2,m_2\rangle=\sqrt{j_1(j_1+1)-m_1(m_1-1)}\nonumber\\
& &\times\sqrt{j_2(j_2+1)-m_2(m_2+1)}\nonumber\\
& &\times\left[j_1(j_1+1)+j_2(j_2+1)-2(m_1-1)(m_2+1)\right]^{k-1}\nonumber\\
& &\times\langle j_1,m_1,j_2,m_2|\hat{\Lambda}_{n-k}^+|j_1,m_1,j_2,m_2\rangle,
\end{eqnarray}

\begin{eqnarray}\label{deux}
& &\langle j_1,m_1,j_2,m_2|\hat{\Lambda}_{n-k}^+\hat{B}_2\hat{A}^{k-1}\hat{B}_1|j_1,m_1,j_2,m_2\rangle =\sqrt{j_1(j_1+1)-m_1(m_1+1)}\nonumber\\
& &\times\sqrt{j_2(j_2+1)-m_2(m_2-1)}\nonumber\\
& &\times\left[j_1(j_1+1)+j_2(j_2+1)-2(m_1+1)(m_2-1)\right]^{k-1}\nonumber\\
& &\times\langle j_1,m_1,j_2,m_2|\hat{\Lambda}_{n-k}^+|j_1,m_1,j_2,m_2\rangle,
\end{eqnarray}

\begin{eqnarray}\label{trois}
& &\langle j_1,m_1,j_2,m_2|\hat{\Lambda}_{n-k}^+\hat{B}_1\hat{A}^{k-1}\hat{B}_1|j_1,m_1,j_2,m_2\rangle =\sqrt{j_1(j_1+1)-m_1(m_1+1)}\nonumber\\
& &\times\sqrt{j_1(j_1+1)-(m_1+1)(m_1+2)}\nonumber\\
& &\times\left[j_1(j_1+1)+j_2(j_2+1)-2(m_1+1)(m_2-1)\right]^{k-1}\nonumber\\
& &\times\langle j_1,m_1,j_2,m_2|\hat{\Lambda}_{n-k}^+|j_1,(m_1+2),j_2,(m_2-2)\rangle
\end{eqnarray}

\noindent and

\begin{eqnarray}\label{quatre}
& &\langle j_1,m_1,j_2,m_2|\hat{\Lambda}_{n-k}^+\hat{B}_2\hat{A}^{k-1}\hat{B}_2|j_1,m_1,j_2,m_2\rangle =\sqrt{j_2(j_2+1)-m_2(m_2+1)}\nonumber\\
& &\times\sqrt{j_2(j_2+1)-(m_2+1)(m_2+2)}\nonumber\\
& &\times\left[j_1(j_1+1)+j_2(j_2+1)-2(m_1-1)(m_2+1)\right]^{k-1}\nonumber\\
& &\times\langle j_1,m_1,j_2,m_2|\hat{\Lambda}_{n-k}^+|j_1,(m_1-2),j_2,(m_2+2)\rangle.
\end{eqnarray}

\noindent As we can see, the sum rule for $\left[\ell(\ell+1)\right]^n$ can be expressed in terms of the sum rule for lower powers (0, 1, 2, $\cdots$, $n-1$) of $\ell(\ell+1)$, but unfortunately it involves also non-diagonal terms (see Eqs. (\ref{trois}) and (\ref{quatre})). The latter can also be expressed in terms of non-diagonal terms for lower powers of $\ell(\ell+1)$, but the corresponding recurrence relation involves in turn diagonal terms. Therefore, we have to solve coupled recurrence relations, involving ``diagonal'' and ``non-diagonal'' sum rules. It is interesting to note, however, that for $\ell(\ell+1)$, $\left[\ell(\ell+1)\right]^2$ and $\left[\ell(\ell+1)\right]^3$, the recurrence involves only diagonal terms of $\Lambda_k^+$ for $k\ge 3$ (Eqs. (\ref{trois}) and (\ref{quatre}) are identically equal to zero because they involve only $\Lambda_0^+$ and $\Lambda_1^+$ which do not contain any $\hat{B}_1$ or $\hat{B}_2$), and therefore the sum rules can be obtained recursively with a single recurrence on the power of $\ell(\ell+1)$.

\end{document}